\documentclass[a4paper,11pt]{article}

\usepackage[utf8]{inputenc}
\usepackage[T1,T2A]{fontenc}
\usepackage{amsmath,amsfonts,amssymb}

 \usepackage{graphicx}
\usepackage{wrapfig}
\usepackage{subfig}
\usepackage{hyperref}
\usepackage[dvipsnames]{xcolor}

\begin{document}

\begin{titlepage}

\vspace{1.5cm}

\begin{center} {\LARGE \bf  On unitarity in  singlet  inflation with a non-minimal coupling to gravity  } \end{center}

\vspace{1cm}

\begin{center}
  {\bf Oleg Lebedev\(^{\,a}\), Yann Mambrini\(^{\,b}\)  
  and Jong-Hyun Yoon\(^{\,b}\)}
\end{center}
  
\begin{center}
  \vspace*{0.15cm}
  \it{\({}^a\)Department of Physics and Helsinki Institute of Physics,\\
  Gustaf H\"allstr\"omin katu 2a, FI-00014 Helsinki, Finland}\\
  \vspace*{0.15cm}
  \it{\({}^b\)Universit\'e Paris-Saclay, CNRS/IN2P3, IJCLab, 91405 Orsay, France
 }
\end{center}
  
\vspace{2.5cm}

\begin{center} {\bf Abstract} \end{center}
 
 \noindent  We study inflationary models based on a non-minimal coupling of a singlet scalar to gravity,
 focussing on the preheating dynamics and the unitarity issues in this regime. If the scalar does not have significant couplings to other fields,
 particle production after inflation  is far less efficient than that in Higgs inflation. 
 As a result, unitarity violation at large non-minimal couplings requires a different treatment. We find that collective effects in inflaton scattering processes during preheating
 make an important impact on the unitarity constraint.  Within effective field theory,  the consequent upper bound on  the non-minimal coupling is of order  a few hundreds.

\end{titlepage}

\tableofcontents

\section{Introduction}

Inflation is one of the cornerstones of modern cosmology \cite{Starobinsky:1980te,Guth:1980zm,Linde:1981mu,Mukhanov:1981xt}. 
An attractive class of inflationary models is based on a non-minimal scalar coupling to curvature  $\xi$ \cite{Chernikov:1968zm,Fakir:1990eg,Buchbinder:1992rb},
\begin{equation}
\Delta {\cal L}=  {1\over 2}\xi  R\, \phi^2 \;,
\end{equation}
where $R$ is the scalar curvature and $\phi$ is the inflaton. A well-known example   is 
 ``Higgs inflation''  \cite{Bezrukov:2007ep}.
An important feature of these models is that they provide an 
 excellent fit 
to the current PLANCK data \cite{Planck:2015fie}. Furthermore, the presence of the coupling $\xi$ is expected on general grounds, e.g. it is generated radiatively, making the framework theoretically well motivated.  

One of the  constraints on the size of the non-minimal coupling to gravity is imposed by unitarity considerations \cite{Burgess:2009ea,Barbon:2009ya}.
 The inflationary models are based on the effective field theory  (EFT) description and thus are 
valid up to a certain energy scale.
After inflation, when the inflaton background has become negligible, the corresponding EFT cut-off is of order \cite{Burgess:2009ea,Barbon:2009ya}
\begin{equation}
\Lambda \sim {M_{\rm Pl}\over \xi} \;,
\label{intro-lambda} 
\end{equation}
while during inflation, the presence of the large  scalar background raises it to
\begin{equation}
\Lambda_{\rm infl} \sim {M_{\rm Pl}} 
\end{equation}
 in the Einstein frame \cite{Bezrukov:2010jz}.
 If $\xi \gg 1$, the effective description may be problematic after inflation bringing consistency of such models in question. 

In particular, after Higgs inflation, a ``violent'' reheating stage takes place, which is driven mainly by the Higgs couplings to the gauge bosons. Much of the energy density of the system is converted into
energetic gauge bosons with momenta close to the Planck scale \cite{Ema:2016dny}. Clearly, this is problematic since the effective field theory cut-off during reheating is far below the Planck scale.    
This phenomenon was explored further in \cite{Ema:2021xhq}.

In our current work, we study the unitarity issues in singlet scalar inflation driven by $\xi \gg 1$, where the singlet does not have significant couplings to other fields, unlike the Higgs. In this case, 
postinflationary particle production is far less efficient than  that  in Higgs inflation and the corresponding unitarity considerations do not apply.
Nevertheless, it is clear that very large $\xi$ would be problematic: due to inflaton quanta production, the energy density after inflation decreases in time slower than the inflaton background does, 
thus violating the bound (\ref{intro-lambda}) eventually. We quantify this statement using lattice simulations. 
More importantly, we evaluate  the impact of inflaton collective effects on scattering amplitudes   during preheating 
and find that these make the unitarity  bound substantially stronger. Finally, we study the effect of a significant non-derivative inflaton coupling to another field. Our main result is that the unitarity bound 
requires $\xi$ to be below a few hundreds, which is important for inflationary model building.

We confine ourselves to unitarity studies in Higgs-like inflationary models in the {\it metric} formulation. Analogous unitarity bounds become  much looser in Palatini formalism \cite{Bauer:2010jg}, although this set-up is not as minimalistic.
The original Higgs inflation model can also be unitarized, i.e. made UV-complete,  for the price of Higgs inflation being a ``mirage'' in the sense that the inflationary dynamics gets dominated by a different field
\cite{Giudice:2010ka,Lebedev:2011aq,Barbon:2015fla,He:2018gyf}.
Other relevant studies of unitarity issues during and after Higgs  inflation   can be found in  \cite{Fumagalli:2016lls,Hamada:2020kuy,Ito:2021ssc,Karananas:2022byw}. In particular,
\cite{Ito:2021ssc} focuses on explicit calculation of the scattering amplitudes, while \cite{Karananas:2022byw} discusses the relevant field redefinitions which make the computations transparent.
Numerical studies of the preheating dynamics in Higgs-like inflation models have   been performed in \cite{Repond:2016sol,vandeVis:2020qcp,Dux:2022kuk,Joana:2022uwc}.

\section{Inflaton background dynamics}

 We start with the inflaton action in the Jordan frame,
 \begin{equation}
{\cal S}= \int d^4 x \sqrt{-g} \left( {1\over 2} M_{\rm Pl}^2 R +  {1\over 2}\xi  R\, \phi^2   - {1\over 2} g^{\mu\nu} \partial_\mu \phi \, \partial_\nu \phi -V
\right)\;,
\label{action}
\end{equation}
where the metric convention is $-+++$ and $\xi >0$.\footnote{The $\xi$ convention differs from that of  our previous work \cite{Lebedev:2022vwf}.} The potential is
\begin{eqnarray} 
&&V=  {1\over 4} \lambda_\phi \phi^4  \;,
 \end{eqnarray}
 where we have neglected the inflaton mass term irrelevant at large field values. In what follows, we will use the Planck units 
 \begin{equation}
  M_{\rm Pl}=1 \;.
  \end{equation}
and consider a large $\xi $ regime,
\begin{equation}
\xi \gg 1\;.
\end{equation}

\subsection{Basics of inflation in the Einstein frame}

The transition to the Einstein frame, where the curvature-dependent term is    ${1\over 2} R$, is accomplished by the metric transformation \cite{Salopek:1988qh}
\begin{equation}
 g_{\mu\nu}^{\rm E} = \Omega \,  g_{\mu\nu} \;,
\end{equation} 
with 
\begin{equation}
  \Omega = 1  + \xi  \phi^2 \;.
\end{equation} 
This eliminates the scalar coupling to curvature for the price of a non-canonical  $\phi$ kinetic term.  
The canonically normalized inflaton  $\chi$ satisfies 
$ {d \chi \over d \phi} = \sqrt{  1 + \xi (1+6 \xi) \phi^2  \over  (1 +\xi \phi^2)^2 } \;,$ 
which is solved by \cite{Garcia-Bellido:2008ycs}
 \begin{equation}
  \, \chi (\phi) = \sqrt{ 1+6 \xi \over \xi} \, \sinh^{-1} \left(    \sqrt{(1+6 \xi) \xi} \, \phi  \right)
 -\sqrt{6 } \,  \sinh^{-1} \left(       {\sqrt{6} \xi    \phi \over \sqrt{1+ \xi \phi^2}}            \right)\;.
 \label{gen-solution}
  \end{equation}
Inflation takes place when  $\xi  \phi^2 \gg 1$, in which case $\chi \simeq {\sqrt{3/2}} \, \ln \xi \phi^2$ and the Einstein frame potential $V^{\rm E} = V/ \Omega^2$ 
has the form 
 \begin{equation}
 V_E(\chi)\simeq  {\lambda_\phi \over 4 \xi^2 }\, \left(     1- e^{- \sqrt{2\over 3} |\chi |} \right)^2 \;.
\end{equation}
It is exponentially close to a flat potential and provides an excellent fit to the  PLANCK data  \cite{Bezrukov:2007ep,Planck:2015fie}.
The energy density during inflation can be approximated by 
$ V^{\rm E} =3H^2 \simeq {\lambda_\phi \over 4 \xi^2 } \;.$
Computing the $\epsilon $ parameter from the above  potential and applying the COBE normalization \cite{Bezrukov:2007ep}, one gets
\begin{equation}
 {\lambda_\phi \over 4 \xi^2 } = 4 \times 10^{-7} \, {1\over N^2} \;,
 \label{cobe}
\end{equation}
where $N$ is the number of $e$-folds of inflation. For $N=60$, $ {\lambda_\phi \over 4 \xi^2 } \simeq 10^{-10}$. This constraint  implies that there is only one independent variable in our analysis, e.g.  $\xi$. 

Inflation ends when $\chi \sim 1$. After that, the potential becomes quadratic and, subsequently, quartic
\cite{Garcia-Bellido:2008ycs}:
\begin{equation}
V_E(\chi) \simeq \left \{
  \begin{tabular}{ccc}
  $ {\lambda_\phi \over 6 \xi^2 } \chi^2 $ & for   & ${1\over 2 \xi } \ll |\chi | \ll  1  ~,$\\
 ${\lambda_\phi \over 4}   \chi^4$  &  for & $|\chi |\ll {1\over 2 \xi } ~.$ 
  \end{tabular}
\right. 
\end{equation}   
In our work, we focus primarily on these regimes.

\subsection{Postinflationary background dynamics in the Jordan frame}

To study the inflaton background dynamics after inflation, it is often convenient to use the Jordan frame. This also applies to lattice simulations of postinflationary dynamics \cite{Figueroa:2021iwm} since the equations of motion
are easier to handle in the Jordan frame. In this section, we analyze how the system behaves in terms of   two field variables: the homogeneous inflaton field $\phi(t)$ and the scale factor $a(t)$.

In  the Friedmann metric
\begin{equation}
ds^2 = -dt^2 + a(t)^2 \, dx^i \,dx^i \;,
\end{equation} 
the scalar curvature has the form
\begin{equation}
R = 6 \left[    {\ddot a \over a} + \left( {\dot a \over a}\right)^2    \right]\;,
\label{R}
\end{equation}
where the dot denotes the time derivative. As usual, the Hubble rate is defined by $H= \dot a /a$. The individual components of the Ricci tensor $R_{\mu\nu }$ can be found, for instance, in   \cite{Figueroa:2021iwm}.
The system satisfies the Einstein equation 
$ R_{\mu\nu}- {1\over 2} g_{\mu\nu} R =   \, T_{\mu\nu} \, ,$ where $T_{\mu\nu} $ is the scalar energy-momentum
 tensor. Due to the non-minimal   coupling to gravity, it contains a curvature-dependent contribution.

 Since there are two field variables in the system, it is sufficient to use two independent equations of motion (EOM).
 The scalar EOM for the homogeneous  background $\phi$ and the Einstein equation $R_{00}-{1\over 2} g_{00} R = T_{00}$
can be written as 
\begin{eqnarray}
&&\ddot \phi + 3 H\, \dot \phi - {1\over a^2 } \nabla^2  \phi - \xi R \phi + {\partial V \over \partial \phi}=0\;, \\
&& 3H^2 = \rho (\phi) \;.
\label{eom-s}
\end{eqnarray}
These equations can be recast in a more convenient form.
Tracing over the indices of the Einstein equation, $R = -  T_{\mu}^\mu \;,$
 and computing $T_{00}(\phi)=\rho (\phi)$ for the inflaton, one finds  \cite{Figueroa:2021iwm,Lebedev:2022vwf}
\begin{eqnarray}
 && R =  \frac{1}{1+(6\xi+1)\xi\,  \phi^2}  \,  \left[     (1+6\xi)  \,\partial^\mu \phi \partial_\mu \phi +4  V    
+6 \xi \, \phi V^\prime_\phi        \right] \;,          \label{R-expr}    \\
&& \rho (\phi)= 
\frac{1}{2}\dot{\phi}^2+
   V(\phi)-
  3\xi H^2  \phi^2 -6\xi H  \phi \dot{\phi} \;,
\label{R}
\end{eqnarray}
accounting for the difference between our definition of $\xi$ and that of \cite{Lebedev:2022vwf}.
Note the unusual presence of the Hubble-dependent terms in the scalar energy density, which appear due to the kinetic scalar-graviton mixing in the Jordan frame (see the discussion in \cite{Lebedev:2022vwf}).

Plugging the expression for $R$ in the scalar EOM, we get for $6\xi \gg 1$,
\begin{equation}
\ddot \phi + 3 H\, \dot \phi   + {6\xi^2 \dot \phi^2 + \lambda_\phi \phi^2 \over 1 +6 \xi^2 \phi^2}\; \phi =0 \;,
\label{eos-2}
\end{equation}
such that $ {6\xi^2 \dot \phi^2 + \lambda_\phi \phi^2 \over 1 +6 \xi^2 \phi^2}$ can be viewed as the effective mass squared. We observe that there is no tachyonic instability and 
the effective mass squared at large $\phi $ is $\lambda_\phi /(6 \xi^2)$.
To solve the scalar EOM, we need an expression for $H$ in terms of $\phi$. Denoting the  usual $\xi=0$ Hubble rate by $H_0$,
\begin{equation}
3H_0^2 = {1\over 2} \dot \phi^2 + V \;,
\end{equation}
one solves the quadratic equation (\ref{eom-s}) to get
\begin{equation}
H = \sqrt{           {H_0^2 \over 1+ \xi \phi^2} +   \left( {\xi \phi \dot \phi      \over 1+ \xi \phi^2 }  \right)^2 } - {\xi \phi \dot \phi      \over 1+ \xi \phi^2 } \;.
\label{H-H0}
\end{equation}
Plugging this into (\ref{eos-2}), we obtain an equation just for $\phi$, which can be solved numerically. 
Note that at large $\phi$, $H=H_0 / (\sqrt{\xi} \phi) $
which is not constant during inflation, unlike the Hubble rate in the Einstein frame.

    \begin{figure}[h!] 
\centering{
\includegraphics[scale=0.38]{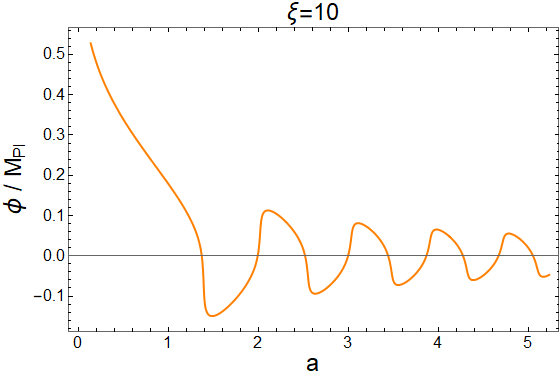}
\includegraphics[scale=0.40]{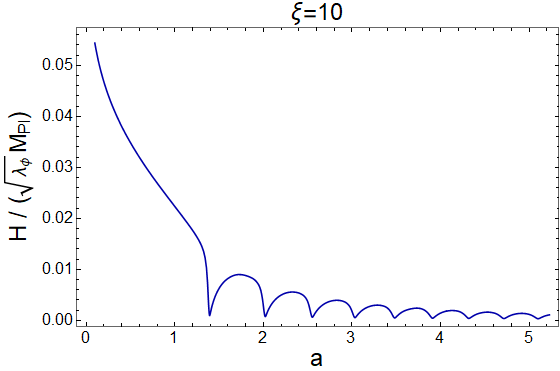}
}
\caption{ \label{phi}
The inflaton and Hubble rate evolution  in the Jordan frame   for $\xi=10$.  The scale factor is normalized to 1 when $R=6H^2$ ($\ddot a=0$) holds true numerically.}
\end{figure}

  The inflationary solution is $\ddot \phi \simeq 0$
 such that $\phi(t)$ is linear, 
 with a small velocity $\dot \phi \sim   \sqrt{\lambda_\phi} \, \xi^{-3/2}$.
 A numerical  example of the field and Hubble rate evolution in the Jordan frame is shown in Fig.\,\ref{phi}.\footnote{
The $\lambda_\phi$ dependence can be eliminated from the scalar EOM by introducing a rescaled time variable $t^\prime =\sqrt{\lambda_\phi} \, t$.} 
 At large field values, both $\phi$ and $H$ are  linear in time. 
 Inflation ends when $\phi$ becomes of order 
 \begin{equation}
 \phi \sim {1\over \sqrt{\xi}}\;,
 \end{equation}
 such that $1+\xi \phi^2$ can no longer be approximated by $\xi \phi^2$. At this point, the velocity contribution  to the effective inflaton mass in (\ref{eos-2}) becomes important,
 $\xi^2 \dot \phi^2 \sim \lambda_\phi \phi^2$. Similarly, the velocity terms in (\ref{H-H0}) start playing a significant role.
  The  energy density at the end of inflation is of order $\rho \sim \lambda_\phi/\xi^2$.
Finally,  when $\phi$ reaches zero, it starts oscillating inducing oscillations in the Hubble rate as well.

 \subsubsection{Spike-like feature in the inflaton evolution}

Ema {\it et al.} \cite{Ema:2016dny} have observed that 
the inflaton field derivative exhibits a spike-like feature  relevant to particle production.
Such a spike is  caused by the shape of the inflaton potential, as viewed in the Einstein frame, because it combines a quadratic potential at larger field values with a quartic potential
at smaller inflaton values.  An earlier study of preheating in this system can be found in \cite{Tsujikawa:1999me}.

Let us examine the appearance of the spike  using our equations of motion. To understand the essence of the mechanism, one may {\it neglect} { the Universe expansion}.
We then have
\begin{equation}
 \ddot \phi + \omega^2 (\phi) \, \phi =0 ~~,~~ \omega^2 = {6\xi^2 \dot \phi^2 + \lambda_\phi \phi^2 \over 1 +6 \xi^2 \phi^2}\;.
 \end{equation}
The quantity $\omega$ has a different behaviour at large and small field values. Consider  first the larger $\phi$ values.

{\bf (a)  \bf{$\xi^2 \phi^2 \gg 1$.}}

\noindent
In this regime, $\omega^2 \simeq \lambda_\phi /(6 \xi^2) + (\dot \phi /\phi)^2$. Let us use  the Ansatz $\phi = \phi_0 \, \cos \omega t$, where $\phi_0$ and $\omega$ are treated  as  
approximately constant.
As long as the field value is large enough, i.e. 
$\cos^2 \omega t \gg \sin^2 \omega t$, the oscillation frequency is indeed almost constant and 
 \begin{equation}
 \omega^2 \simeq {\lambda_\phi \over 6 \xi^2} \;.
 \end{equation}
This expression therefore gives the frequency of large-amplitude oscillations.

    \begin{figure}[h!] 
\centering{
\includegraphics[scale=0.399]{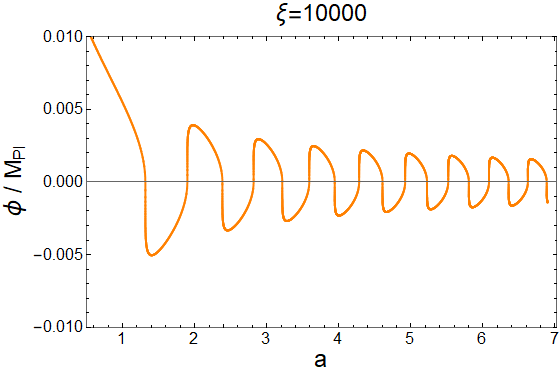}
\includegraphics[scale=0.403]{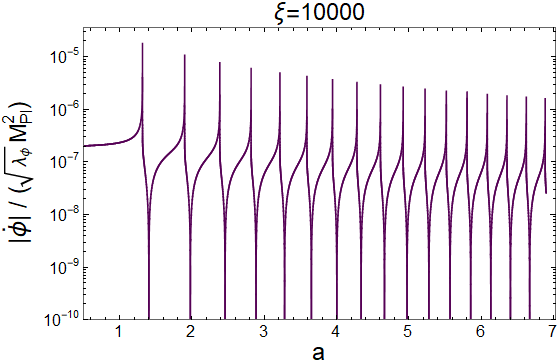}
}
\caption{ \label{phi-dot}
The inflaton field and its velocity evolution in the Jordan frame at $\xi=10^4$. The scale factor is normalized to 1 when $R=6H^2$ ($\ddot a=0$) holds true numerically.}
\end{figure}

{\bf (b)  \bf{$\xi^2 \phi^2 \ll 1$.}}

\noindent
Consider the small field limit $\phi \rightarrow 0$. In this case,
\begin{equation}
\omega^2 \simeq 6\xi^2 \dot \phi^2
\label{omega-1}
\end{equation}
 and $\omega$ cannot {\it a priori} be treated as a slowly varying function.
The EOM becomes 
\begin{equation}
 \ddot \phi +   6\xi^2 \dot \phi^2\, \phi =0  \;.
 \end{equation}
This is easily solved for $\dot \phi$,
\begin{equation}
\dot \phi = {\rm const} \times e^{-3\xi^2 \phi^2} \;,
\label{dot-phi}
 \end{equation}
where the constant is determined by the boundary conditions. In the small field regime, the exponent is close to one, so $\dot \phi \simeq  {\rm const} $.
The size of the constant can be evaluated using approximate energy conservation. The energy density at the end of inflation, $\lambda_\phi/\xi^2$, gets converted into
the kinetic energy at $\phi =0$, so $\dot \phi^2 \sim \lambda_\phi/\xi^2$.
 Hence  
\begin{equation}
 \dot \phi \sim \sqrt{\lambda_\phi} /\xi 
 \end{equation}
 and 
  \begin{equation}
  \omega \sim    \sqrt{\lambda_\phi}  \, .
 \end{equation}
This corresponds to a much sharper variation  of $\phi$ compared to the large field case, by a factor of $\xi$.
As $|\phi|$ increases away from  the zero crossing, the velocity becomes exponentially suppressed according to (\ref{dot-phi}), which generates a spike in $\dot \phi$.

The Hubble friction does not change this behaviour fundamentally, so the full system exhibits similar features.
The main point is that the frequency $\omega$ characterizing the variation of $\phi$ in time changes abruptly around inflaton zero crossings $\phi =0$, which indicates non-adiabaticity.
A numerical example illustrating the presence of spike-like features in shown in  Fig.\,\ref{phi-dot}. Due to the logarithmic scale, there also appear ``negative'' spikes corresponding to $\dot\phi \rightarrow 0$.

\subsubsection{Non-adiabaticity}
\label{sec-ad}

Semiclassical particle production is characterized by the non-adiabaticity parameter $\dot \omega / \omega^2$. The oscillation frequency $\omega$ changes abruptly at the border of the regions considered above,
so let us focus on $ \xi |\phi | \lesssim 1$. In this case, 
\begin{equation}
\omega \simeq { \sqrt{6} \xi |\dot \phi |   \over \sqrt{1 +6 \xi^2 \phi^2}} \;.
\end{equation}
This is because  $\lambda_\phi \phi^2 \ll \xi^2 \dot \phi^2$ in this region. Indeed, approximate energy conservation requires 
$\lambda_\phi /\xi^2 \sim \dot \phi^2 + \lambda_\phi \phi^4 $, while the potential term is negligible at $\phi \sim 1/\xi$.

    \begin{figure}[h!] 
\centering{
\includegraphics[scale=0.40]{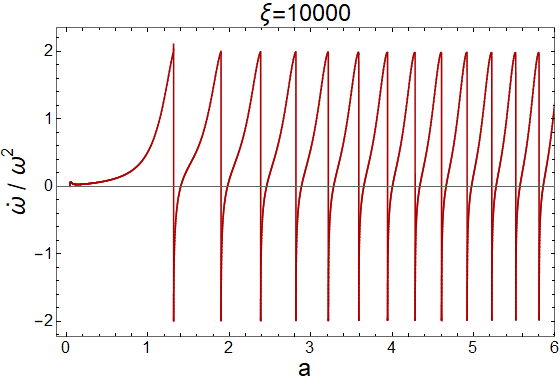}
}
\caption{ \label{non-ad}
The non-adiabaticity parameter for $\xi =10^4$. The scale factor is normalized to 1 when $R=6H^2$ ($\ddot a=0$) holds true numerically.}
\end{figure}

One then finds for $\dot \phi >0$,
\begin{equation}
 {\dot \omega \over  \omega^2} \simeq -2 \,  { \sqrt{6} \xi  \phi    \over \sqrt{1 +6 \xi^2 \phi^2}} \;.
\end{equation}
The second factor in this expression  is bounded by one, so $ | \dot \omega / \omega^2 | > 1$ at $ \xi \phi \sim 1$ with the maximal value of 2. 
This behaviour is seen in Fig.\,\ref{non-ad}.

These considerations can be extended to modes with non-zero momenta by replacing
\begin{equation}
\omega^2 \rightarrow {k^2 \over a^2} + \omega^2 \;.
\end{equation} 
As long as $k^2/a^2 \lesssim \omega^2$, the above conclusions apply. In the non-adiabatic region, $\omega \lesssim \sqrt{\lambda_\phi}$.
Therefore, the momentum cut-off for the produced particles is of order
\begin{equation}
{k\over a} \lesssim \sqrt{\lambda_\phi} \;.
\end{equation}

\section{Postinflationary dynamics on the lattice and unitarity}

The dynamics of the inflaton field after inflation are complicated by backreaction and rescattering effects. These lead to fast decay of the zero mode and growth of inhomogeneities, both of which have an impact on unitarity violation considerations.
To take such effects into account, we resort to lattice simulations using the tool CosmoLattice \cite{Figueroa:2020rrl,Figueroa:2021yhd}.

    \begin{figure}[h!] 
\centering{
\includegraphics[scale=0.31]{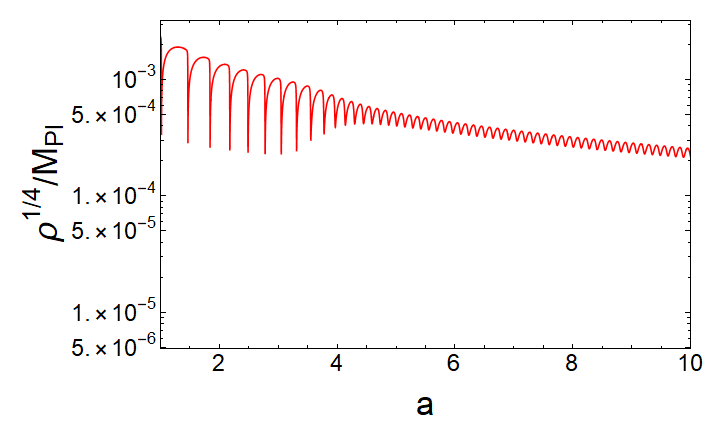}
\includegraphics[scale=0.31]{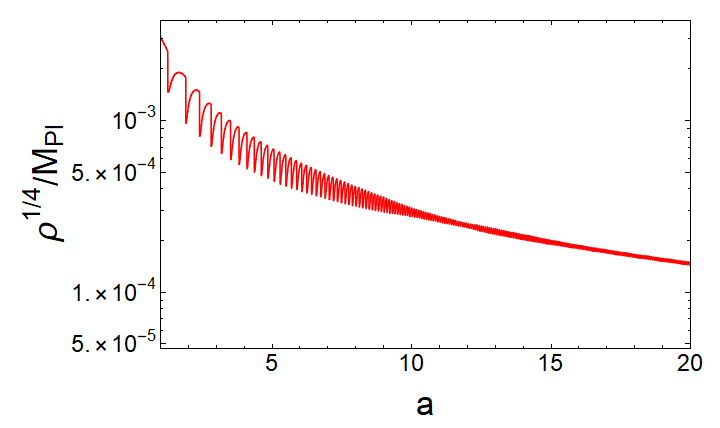}
}
\caption{ \label{energy}
Evolution of the energy density scale $\rho^{1/4}$ in Planck units. {\it Left:} $\xi=100$, $\lambda_\phi =4.4 \times 10^{-6} $.
{\it Right:} $\xi=2000$, $\lambda_\phi =1.8 \times 10^{-3} $. }
\end{figure}

One of the features of CosmoLattice is a rearrangement of the  second-order equations
into the first-order ones. 
However, in the presence of the non-minimal coupling,  this becomes problematic since the 
pressure and
energy density terms in the second-order Friedmann equation have an explicit dependence on
the scale factor. Instead, we use a customized version of CosmoLattice as in  \cite{Figueroa:2021iwm}, which solves for the scale 
factor with the help of  Eq.\,\ref{R-expr} and an explicit Runge-Kutta   algorithm. 
We employ a lattice with 320 sites along each axis, which is sufficient to capture the 
characteristic power spectra. We observe good convergence of the results and consistency with other studies such as 
   \cite{Figueroa:2021iwm}, when applicable.

The initial conditions for the simulation are chosen as follows: at the end of inflation, $\phi\sim 1/\sqrt{\xi}$ in Planck units, we neglect $\dot \phi$  and solve Eq.\,\ref{eos-2} to find $\phi,\dot \phi$ at later 
times.\footnote{We find that the results are insensitive to the initial value of $\dot \phi$ as long as it is small.} 
At the same time, we evolve the field fluctuations  with a Python code to set up the boundary conditions for the CosmoLattice simulations.
For the latter, we use $\phi=0.01 $ as the  initial value ($a=1$) and adopt the normalization (\ref{cobe}) with $N=60$.
This implies, for example, that at $a=1$,
$\rho^{1/4} = 0.0023\, $ for $\xi=100$ and 
$\rho^{1/4} = 0.00247\,$ for $\xi=500$.
At this stage, we set vacuum fluctuations as the initial conditions for the inflaton momentum modes and evolve these with the help of  lattice simulations. We find that using inflationary fluctuations instead
does not make a visible difference since inflation affects superhorizon modes $k/a \lesssim H$, while the hard modes are much more important for our purposes (see also  \cite{Figueroa:2021iwm}).

 Given the  shape of the inflaton potential,  one expects the energy density to scale as non-relativistic matter initially, 
  while at later times, the scaling  becomes radiation-like.
  In practice, the field fluctuations play an important role such that 
  the truly non-relativistic regime shortens significantly  and most of the time at $\phi > 1/\xi$   the system behaves as semirelativistic, i.e. exhibits a scaling law between $a^{-3}$ and  $a^{-4}$.
 For $ \phi \lesssim 1/\xi $, the system becomes radiation-like, so 
 \begin{eqnarray}
 &&\rho \propto a^{-(3+x)} ~~~,~~~ 1/\xi \lesssim \phi \lesssim 1/\sqrt{\xi}  ~~, \nonumber \\
  &&\rho \propto a^{-4}  ~~~,~~~ \phi \lesssim 1/\xi  ~~,
 \end{eqnarray}
 where $0<x<1$.
 Numerical examples of the energy density evolution are shown in Fig.\,\ref{energy}. The apparent strong oscillatory behaviour is due to lattice limitations, yet the above scaling applies to $\rho$ averaged over a few oscillation periods.

    \begin{figure}[h!] 
\centering{
\includegraphics[scale=0.31]{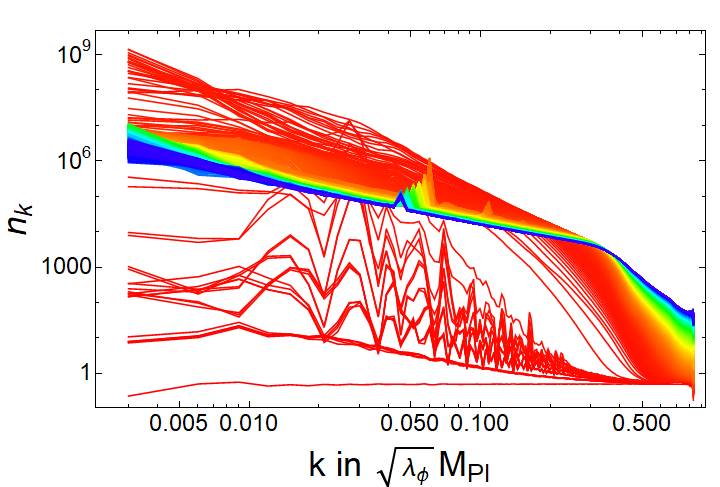}
\includegraphics[scale=0.31]{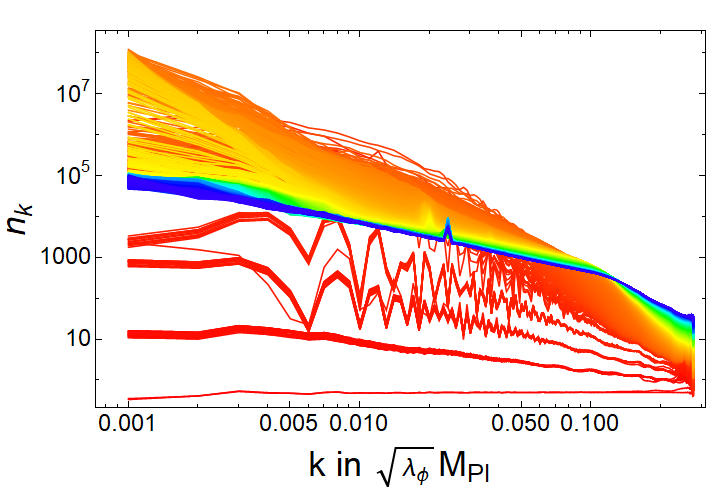}
}
\caption{ \label{specrtum}
Evolution of the inflaton momentum spectrum (red to blue). 
{\it Left:} $\xi=100$, $a_{\rm end}=300$.
{\it Right:} $\xi=500$, $a_{\rm end}=70$.  $n_k$ is the occupation number and the  comoving momentum $k$  is in units of $\sqrt{\lambda_\phi}$,
where $\lambda_\phi$ is fixed by the COBE normalization.}
\end{figure}

The oscillating inflaton background evolves non-adiabatically (Sec.\,\ref{sec-ad}) and 
resonantly excites non-zero momentum modes via an analog of Eq.\,\ref{eos-2}.\footnote{The EOM for the different momentum modes do not decouple and cannot be written in a simple way, e.g. due to momentum-dependent contributions to the curvature. Instead, one solves the system on the lattice in coordinate space as detailed in  \cite{Figueroa:2021iwm,Lebedev:2022vwf} and then decomposes the solution in the Fourier modes.}
  The process is highly non-linear 	and 
can be studied reliably only on the lattice.
Expanding $\phi(x)$ in 3-momentum modes  $\phi_k$ as in \cite{Greene:1997fu}, it is convenient to define a rescaled variable $Y_k \equiv a \, \phi_k$ as a function of conformal time $\tau$: 
$d\tau =d t/a$.  Then, the oscillator number  $n_k$ and the corresponding frequency $\omega_k$
are defined by
\begin{eqnarray}
&& n_k \equiv \frac{1}{2} \left( \omega_k |Y_k|^2 +\frac{1}{\omega_k} |\dot Y_k |^2 \right) \;, \\
&& \omega^2_k \equiv k^2+a^2 \left\langle \frac{\partial^2 V(\phi)}{\partial \phi^2} \right\rangle \;,
\end{eqnarray}
where $\langle ...\rangle$ denotes spacial averaging.
$n_k$ changes in time signifying particle production and rescattering.
Evolution of the (comoving) momentum spectrum for two representative examples $\xi=100$ and $\xi= 500$ is shown in Fig.\,\ref{specrtum}. One observes the momentum cut-off
of order $\sqrt{\lambda_\phi}$, as expected. The typical occupation numbers are very large, $n_k \gg 1$, manifesting  the importance of collective effects. At late times, 
one also observes an isolated peak in the spectrum. It is due to the momentum modes excited in the $\phi^4$ potential via the Lam\'e equation \cite{Greene:1997fu}. The magnitude of
this momentum band is given approximately by 
\begin{equation}
k \sim \sqrt{\lambda_\phi} \,  \phi_* \;,
\end{equation}
where $\phi_* \sim 1/\xi$ corresponds to the onset of the quartic  potential regime. This only gives a rough estimate since 
the velocity of the field at $ \phi_*$ is non-zero, unlike in the analysis of \cite{Greene:1997fu}, and $\phi_*$ is not precisely defined. Finally, 
 the bump is expected to be smoothed out on a larger time scale due to rescattering.

\begin{figure}[h!] 
\centering{
\includegraphics[scale=0.32]{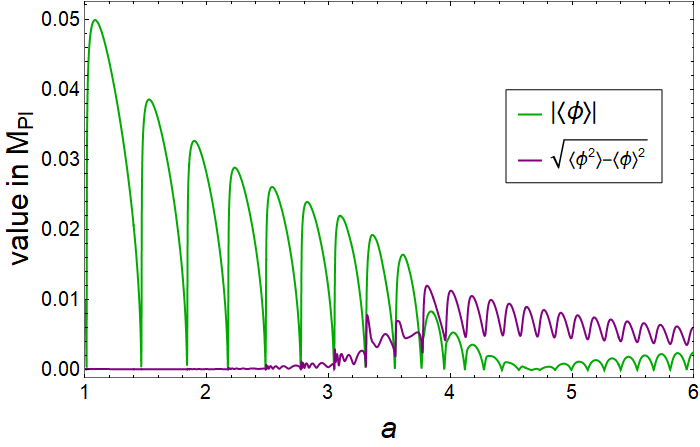}
\includegraphics[scale=0.32]{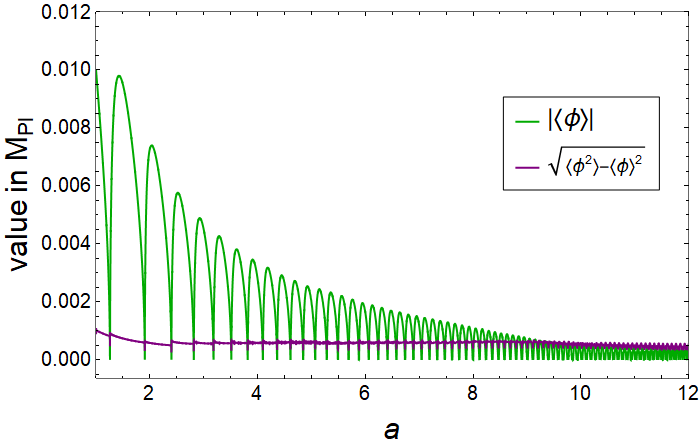}
}
\caption{ \label{phi-evol}
Evolution of the inflaton zero mode and the inflaton variance. {\it Left:} $\xi=100$, $\lambda_\phi =4.4 \times 10^{-6} $.
{\it Right:} $\xi=2000$, $\lambda_\phi =1.8 \times 10^{-3} $. }
\end{figure}

An important ingredient in our study is the evolution of the inflaton zero mode. The inflaton background affects the unitarity bound, hence it is essential to determine 
its decay time.  Fig.\,\ref{phi-evol} displays the evolution of the background $|\langle \phi  \rangle |$ and the fluctuation amplitude $\sqrt{\langle \phi^2  \rangle -\langle \phi  \rangle^2 }$
for $\xi=100,\,2000$.
We observe that at large $\xi$, the fluctuations become larger than the background when the zero mode is of the size
\begin{equation}
\phi_* \sim {1\over \xi} \;.
\end{equation}
This inflaton value signifies the transition to the quartic potential regime. From this point on, the system becomes fully dominated by fluctuations and the background can be neglected for most purposes.
 (For small $\xi\sim 5$, these conclusions do not apply and the fluctuations are much suppressed.)

 \subsection{Unitarity violation: simple estimate}

 When the inflaton background value is negligible,  a ballpark estimate of the  unitarity bound on $\xi$ can be obtained  
 by analyzing the 
  energy density of the system. Since $ \xi \langle \phi^2 \rangle \ll 1$, the difference between the Jordan and Einstein frames is insignificant in terms
  of the energy scales, hence one can use either frame to estimate the unitarity bound.
  
 The canonically normalized inflaton variable $\chi$ in the $Einstein$ frame 
    is defined by the differential equation
 \begin{equation}
{d \chi \over d \phi} = \sqrt{  1 + \xi_\phi (1+6 \xi_\phi) \phi^2  \over  (1 +\xi_\phi \phi^2)^2 } \;.
\label{dchi/dphi}
 \end{equation}
 The solution is given by (\ref{gen-solution}), 
although it is often more convenient to work directly with (\ref{dchi/dphi}) in specific limits.

  Since the kinetic and the curvature terms are canonical, all the non-trivial physics resides in the scalar potential $V= \lambda_\phi \phi^4 / \left[ 4(1+ \xi \phi^2)^2 \right]$ formulated in terms of $\chi$.
Given the inflaton background value at the particular evolution stage, one expands the potential in terms of fluctuations over the background. 
Perturbative unitarity within the 
effective field theory description 
then imposes a constraint on the energy scale of the fluctuations.
During inflation, the expansion is of the form $(\delta \chi)^n$ so that the unitarity violation scale is $M_{\rm Pl}$ \cite{Bezrukov:2010jz}. The same bound  also applies at the end of inflation when $\chi \sim \xi \phi^2 \sim 1$.
This can be shown by examining  (\ref{dchi/dphi}) in the vicinity of $\xi \phi^2 \sim 1$.  However, at very small $\phi \ll 1/\xi$ and large $\xi \gg 1$, 
\begin{equation}
\phi \simeq \chi \, (1-\xi^2 \chi^2) \;.
\label{phi-chi}
\end{equation}
The  expansion around the vacuum then has the form $(\xi \chi)^n$, which implies that the perturbative  scattering amplitudes based on $V(\chi)$  blow up at the energy scale $1/\xi$ .\footnote{The size of the coupling $\lambda_\phi$ does not matter for  a sufficiently large $n$.}   
The characteristic energy scale for the scattering processes can be taken to be $\rho^{1/4}$  at the time when the background becomes negligible, so the unitarity limit corresponds to
\begin{equation}
\rho^{1/4} (\phi_*) \sim {1\over \xi} \;.
\label{bound-1}
\end{equation}
The resulting bound on $\xi $ can be estimated by  $\left({\lambda_\phi / 4\xi^2}\right)^{1/4} \times  {a_*^{-3/4}}  \sim {1/ \xi} $, where $a_*$ is the scale factor corresponding to the decay of the zero mode
and we have assumed non-relativistic scaling of the energy density before $a_*$.  Since typically $a_*$ is between 5 and 10, this gives the critical value of $\xi$ around 1000.
A more careful numerical analysis yields
\begin{equation}
\xi_{\rm max}^0 \simeq 2000 \;.
\end{equation}  
  Here the superscript $0$
  serves to emphasise that this result is based on simple dimensional analysis.

 \subsection{Impact of collective effects on the unitarity bound}

 After inflation, the inflaton field exhibits complicated dynamics where collective effects are important. The system can be treated as a collection inflaton quanta with given occupation numbers $n_k$,
 which can be very large. 
 On the other hand,
 the typical momentum can be much below the naive estimate $\rho^{1/4}$. In what follows, we estimate the impact of collective effects on the unitarity bound within the effective field theory approach.

To be specific, let us study the Higgs quanta production by the inflaton field. Since the system must be reheated at some stage,  consider a small trilinear coupling in the Jordan frame,
\begin{equation}
\Delta V = {\sigma_{\phi h}} \, \phi \, H^\dagger H \;,
\end{equation}
 which leads to inflaton decay at late times. In the Einstein frame, the corresponding interaction is $\Delta V / \left[ (1+ \xi \phi^2)^2 \right]$ written in terms of $\chi \ll 1$. 
 We  may choose  ${\sigma_{\phi h}}$ to be so small that it does not affect the preheating dynamics.
 When the inflaton background becomes negligible, one 
 can use (\ref{phi-chi}) or,  more precisely, a high order  expansion of (\ref{gen-solution}) around zero, which gives $\phi$ as a series in $(\xi \chi)^n$ plus subleading terms. As a result, 
 we obtain a set of interactions
  \begin{equation}
 {\sigma_{\phi h}} \, H^\dagger H   \; {\chi^n \over \Lambda^{n-1}} ~,
 \label{eff-int}
\end{equation}
where 
\begin{equation}
\Lambda \equiv {1\over \xi} \;.
\end{equation}
 Such interactions induce, in particular, $n \rightarrow 2$  Higgs production processes. Their efficiency depends on the mode occupation numbers or, in other words, momentum distribution function.
 For highly occupied modes, the corresponding amplitude can become large violating perturbative unitarity.
 
 To make simple estimates, let us approximate the inflaton momentum distribution function by a step function,
  \begin{equation}
  f(p) = f \times \theta (p_{\rm max}- |p|) \;,
 \end{equation}
 where $f$ is a constant and $p_{\rm max}$ corresponds to the ``maximal'' 3-momentum.  The inflaton number density and the energy density are defined by 
 $ n = \int {d^3p \over (2\pi)^3} \, f(p)$ and 
  $ \rho =        \int {d^3p \over (2\pi)^3} \, E_p \,f(p)    \;,$ respectively. In practice, the inflaton quanta can be treated as relativistic, $E_p \simeq |p|$ since the 
  induced inflaton mass $\sqrt{\lambda_\phi \langle \phi^2 \rangle} $  is smaller than the typical momenta.  
  Then, 
  \begin{equation}
  n =  {1\over 6\pi^2}\; p_{\rm max}^3 \,f  ~~,~~\rho =   {1\over 8\pi^2} \; p_{\rm max}^4  \,f \;.
  \label{nrho}
  \end{equation}
  We observe that for  $f\gg 1$, the characteristic momentum $p_{\rm max}$ is far below $\rho^{1/4}$.

The probability of the Higgs pair production with fixed momenta via (\ref{eff-int}) is given by
\begin{equation}
\rho (n\rightarrow 2) = {1\over n !}  \int \left( \prod_{i=1}^n  {d^3 k_i \over  (2\pi)^3 2E_i}  \,f(k_i)\, \right)  (2\pi)^4 \delta^{(4)} \left(\sum_i K_i -\sum_j P_j \right) \; |{\cal M}(n\rightarrow 2)|^2 \;,
\end{equation}
 where $K_i$ and $P_j$ are the initial and final 4-momenta, respectively; ${\cal M}(n\rightarrow 2)$ is the standard QFT transition amplitude, and $n!$ accounts for identical particles in the initial state.
 Let us normalize the coefficient of operator (\ref{eff-int}) to $1/n!$ to avoid large combinatorial factors in the amplitude. Then,
 performing the integrals and using the Stirling approximation for $n!$, one finds that for sufficiently large $n$, the transition probability grows as 
  \begin{equation}
\rho (n\rightarrow 2) \propto  \left(   c_n \, {p_{\rm max} \sqrt{f} \over \Lambda}    \right)^{2n}  \;,
\label{probability}
\end{equation}
with $c_n = \sqrt{e/(8 \pi^2 n)}$.  The prefactor in  (\ref{probability}) is unimportant for large enough $n$ so that the precise value of $\sigma_{\phi h }$ does not play any role.
If the factor in the parentheses is significantly larger than one, the probability grows uncontrollably violating unitarity. We note that the precise value of $c_n$ depends on the coefficient
of operator (\ref{eff-int}): for instance, if it is order 1 instead of $1/n!$, the $\sqrt{n}$ factor in $c_n$ would appear in the numerator instead of the denominator. Given this ambiguity, in what follows we 
will simply assume $c_n \sim {\cal O}(1)$.\footnote{Although the coefficients of higher dimensional operators can be computed, their signs alternate. Given that our system is a superposition of states with fixed particle numbers (``squeezed'' state), the collective impact  of these operators  is difficult to estimate, hence we simply focus on a single term assuming $c_n \sim 1$. 
 }

We conclude that the figure of merit for the effective field theory expansion is roughly 
\begin{equation}
 \kappa \sim {p_{\rm max} \sqrt{f} \over  \Lambda}  
\end{equation}
for moderately large $n$, and unitarity requires
\begin{equation}
\kappa \lesssim 1\;.
\end{equation}
Hence, highly occupied states can cause problems   even for low characteristic energy of the inflaton quanta.
To give an example, we find that for $\xi =500$ at $a\sim 6$ corresponding to the decay of the inflaton zero mode,\footnote{We obtain these numbers from the lattice output
by parametrizing the  physical   energy  and particle densities    in terms of $p_{\rm max} $ and $f$. These quantities are obtained in the Jordan frame, however, at  this stage $\phi \sim \chi$ and  the difference between the frames 
is insignificant for our purposes.}
\begin{equation}
p_{\rm max} \sim 10^{-2} \sqrt{\lambda_\phi} ~~,~~ f \sim 10^4 \;,
\end{equation}
which makes $\kappa \sim 5$. The inflaton effective mass at this stage is of order $\sqrt{\lambda_\phi  \langle \phi^2 \rangle} \sim 
\sqrt{\lambda_\phi} /\xi   $, which is below $p_{\rm max}$ justifying our relativistic approximation.
For larger $\xi$, $\kappa$ grows further: the main factors are  
 $p_{\rm max} \propto \sqrt{\lambda_\phi} \propto \xi$  due to the inflationary constraint and $1/ \Lambda \propto \xi$, making 
$\kappa \propto \xi^2$. Even though $\sqrt{f}$ decreases tending to ${\cal O}(1)$ at very  large $\xi $
and the decay time for the zero mode  increases slowly with $\xi$ (Fig.\,\ref{phi-evol}), these factors 
do not overcome the $\xi^2$ growth. 
We  conclude that unitarity is violated at
\begin{equation}
\xi_{\rm max} \sim {\rm few}\, \times \, 100 \;
\label{final-bound}
\end{equation}
according to the above criterion, keeping in mind the uncertainty associated with $c_n$.

We see that collective effects lead to a stronger unitarity bound. This tendency can be understood as follows.
 If we impose a bound on the energy density of the system, this constrains $p_{\rm max} \,f^{1/4}$  
according to  (\ref{nrho}). On the other hand, the combination that appears in scattering processes is $p_{\rm max} \,f^{1/2}$. Thus, we have 
 \begin{equation}
    p_{\rm max} \,f^{1/2}  \gg    p_{\rm max} \,f^{1/4} \,    ,
    \end{equation}
and  the scattering bound is stronger 
 by the factor $f^{1/4 } \gg 1 $.
 For small occupation numbers, on the other hand, the results are similar. Below we  illustrate the above statement with a simple example.

{\bf {\underline{Example}}.}  Let us apply  an analogous analysis to Higgs production by an inflaton background in the $\phi^2$ potential. The process  can be viewed as annihilation of the  non-relativistic inflaton quanta.
 Suppose $V={1\over 2} m^2 \phi^2$ and the interaction term is $ \phi^4 h^2/  \Lambda^2$. The momentum distribution function for the inflaton  can be written as
 $f(p) = (2\pi)^3 \delta^{(3) } (p)\,n$, where $n$ is the number density. 
 Although this system is non-relativistic, $n$ and $\rho$ can be expressed in terms of the parameters of  our box-shaped distribution function.  
 Since $\rho= mn$, one finds $p_{\rm max} = 4m/3$ which plays the role of the characteristic energy scale. 
 The average occupation number is then found via $f=8\pi^2 \,\rho/ p_{\rm max}^4$, which yields 
   $f\sim 15 \, \phi^2/m^2 \gg 1$. 
  In the case at hand, we can also take into account the effect of $c_n$: with the above normalization, the analog of $c_n$ is about $1/7$ such that 
  $c_n \; {p_{\rm max} \sqrt{f}} /\Lambda $ becomes unity at 
  $\phi \sim \Lambda$, which signals unitarity violation. Therefore, we obtain the expected result. The cutoff 
 $\Lambda$ is far above the typical energy of the inflaton quantum $m$ and the large occupation number plays a crucial role.  
 Note that requiring $\rho^{1/4}$ of the inflaton field to be below the cut-off would impose a {\it very different} constraint $\phi < \Lambda^2 /m$.

Similar considerations apply to inflaton self-interaction $\chi^{l+4}/ \Lambda^l$. The corresponding $n \rightarrow 2$  reaction rate is enhanced by a further factor $f^2$ from the final state,  which is however inconsequential for 
a sufficiently large $l$. Thus, one again obtains the unitarity bound (\ref{final-bound}). Note also that, if the bound is violated, the system becomes sensitive to arbitrarily small couplings to new fields, which signals
pathology, at least within effective field theory.

It is important to understand the limitations of our analysis. We have relied on perturbative description 
  within {\it effective field theory}, yet violation of perturbative unitarity does not always imply fundamental problems \cite{Dvali:2010jz,Dvali:2022vzz}. 
  Also, we have dealt with $n$-particle states, whereas the system in reality is a complicated superposition of such states. However,
  we believe that our analysis captures the main physical features and results in a reasonable estimate of $\xi_{\rm max}$, above which the effective field theory
  description becomes inadequate.

 \subsection{Effect of a significant Higgs-inflaton coupling}

    \begin{figure}[h!] 
\centering{
\includegraphics[scale=0.36]{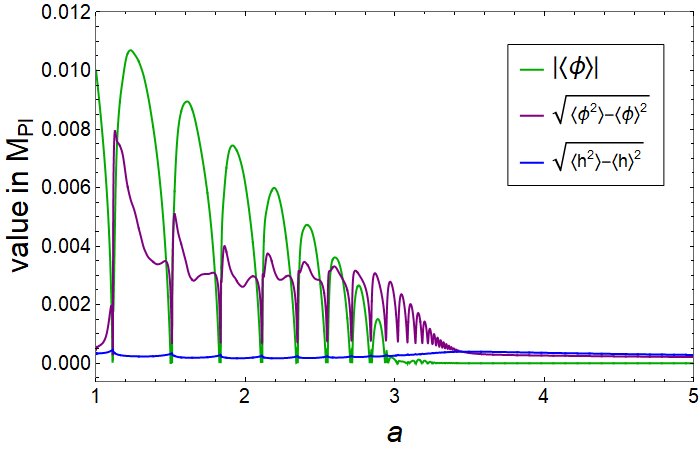}
}
\caption{ \label{lh0.5}
 Evolution of the inflaton zero mode and the variances in the presence of the inflaton-Higgs coupling $\lambda_{\phi h}=0.5$ at
 $\xi=1000$ with 4 Higgs d.o.f.
}
\end{figure}

A significant inflaton coupling to other fields, e.g. the Higgs boson \cite{Lebedev:2021xey,Gross:2015bea}, leads to more efficient particle production and quicker background decay.
Fig.\,\ref{lh0.5} shows an example for  $\xi=1000$ with $\lambda_{\phi h}=0.5$, where the  coupling is defined by
 \begin{equation}
\Delta V =  {1\over 2} {\lambda_{\phi h}} \, \phi^2 \, H^\dagger H \;.
\end{equation}
The zero mode decays by $a\sim 3$, which makes its lifetime shorter by a factor of 2-3 compared to the zero coupling case.
 This occurs due to strong backreaction effects of the produced Higgses on the inflaton background. 
 By the time $a\sim4$, the Higgs field variance becomes larger than that of the inflaton. At this stage, the system reaches quasiequilibrium such that each degree of freedom carries the same fraction of the total energy.
 Since the Higgs field has 4 d.o.f. at high energies, it dominates. 
 We find that the Higgs self-coupling   
between 0 and $10^{-2}$ at this  scale does not affect the results in any significant way.
We also note that the presence of a non-minimal Higgs coupling to gravity  $H^\dagger H R$  would  accelerate  the 
decay of the inflaton zero mode, yet its effect can largely be captured by increasing the Higgs-inflaton coupling.

Since the background decays very quickly, the simple unitarity bound based on $\rho^{1/4} (\phi_*) \sim {1/\xi} $ becomes stronger. For the above parameters, it requires 
 $\xi < 1000$. Another effect of a strong inflaton-Higgs coupling is that the spectrum shifts to the UV  which increases $\kappa$. For instance, $\xi=1000$ yields $\kappa \sim 20$ 
 at the background decay time. The resulting $\xi_{\rm max} $ decreases by  about a factor of 2. Although this makes the unitarity constraint stronger, the improvement does not change
 the estimate (\ref{final-bound}) fundamentally, given the uncertainties involved.

  A significant Higgs-inflaton coupling is interesting in the context of inflaton dark matter \cite{Lerner:2009xg}  and inflaton thermalization  \cite{Lebedev:2021ixj}. As shown recently in \cite{Lebedev:2021zdh},
  the minimal inflaton dark matter model is only viable if $m_\phi \simeq m_h/2$. This conclusion was based on the assumption that the non-minimal coupling $\xi$ cannot exceed about 300 without
  violating unitarity. Our current work supports this assumption and reinforces the conclusions of     \cite{Lebedev:2021zdh}.  We note also that 
  the direct dark matter detection limits have become significantly  stronger \cite{LZ:2022ufs},  which puts many variants of Higgs portal dark matter under extra pressure
  and confines inflaton dark matter to a narrow resonance region.

  Particle production becomes even more efficient if the inflaton has derivative couplings, as in original Higgs inflation \cite{Bezrukov:2007ep}. This case was studied in \cite{Ema:2016dny} with the conclusion 
  that, for $\xi \gg 1$, much
  of the inflaton energy can be converted into particles already within a single inflaton oscillation. Clearly, this is problematic for unitarity since Higgs inflation requires $\xi \sim 5\times 10^4$.

  Finally, let us note that this framework offers interesting reheating, dark matter  and leptogenesis phenomenology \cite{Clery:2022wib,Co:2022bgh,Barman:2022qgt}.
   It does not require an explicit coupling of dark matter to the inflaton. Since the inflationary scale is high, dark matter is abundantly produced by gravitational effects 
  during inflation and/or preheating \cite{Lebedev:2022cic}.

  \section{Conclusion}

 Models of inflation driven by 
  a non-minimal scalar coupling to curvature  are among the most attractive and experimentally viable models of the Early Universe.
  We have studied postinflationary dynamics in a singlet scalar model of this type. 
  In particular, employing lattice simulations,
  we have focused on perturbative unitarity constraints after inflation with a large non-minimal coupling $\xi$.  
  The main ingredients in our study are the decay time of the inflaton zero mode and collective effects associated with large
  occupation numbers of the  inflaton momentum modes. 
  We find that for values of $\xi$ above a few hundreds, 
   such collective effects lead to large scattering amplitudes within effective field theory, thereby violating perturbative unitarity.   
   This is the case even if the inflaton does not have any substantial  couplings to other fields.  We thus obtain  an upper bound on $\xi$, at least, in the effective field theory description.
   If the inflaton has a significant  coupling to other fields, particle production becomes more efficient and the unitarity bound becomes stronger. 
       \\ \ \\
       {\bf Acknowledgements.} We are grateful to the authors of \cite{Figueroa:2021iwm} for providing us with a customized version of CosmoLattice.
   This work was performed using HPC resources from the ``M\'esocentre'' computing center of CentraleSup\'elec, \'Ecole Normale Sup\'erieure Paris-Saclay and Universit\'e Paris-Saclay supported by CNRS and R\'egion \^Ile-de-France (https://mesocentre.universite-paris-scalay.fr/).     
 This project has received support from the European Union's Horizon 2020 research and innovation programme under the Marie Sklodowska-Curie grant agreement No 860881-HIDDeN, and the IN2P3 Master Project UCMN.

\end{document}